\documentclass[conference]{sig-alternate-2013}
\usepackage{epsfig,booktabs}
\usepackage{amsfonts,amsmath}
\usepackage{float}
\usepackage{microtype}
\usepackage{comment}
\usepackage{footnote}
\usepackage[group-separator={,}]{siunitx}
\usepackage[export]{adjustbox}

\usepackage[table]{xcolor}
\usepackage{color, colortbl} \definecolor{linkcol}{rgb}{0,0,1}
\definecolor{citecol}{rgb}{0,0,1}
\definecolor{urlcol}{rgb}{0,0,1}

\usepackage[hyphens]{url} \usepackage[bookmarks=true,bookmarksnumbered=true,linkcolor=linkcol,citecolor=citecol,urlcolor=urlcol,hypertexnames=true,pdfpagelabels]{hyperref}
\usepackage{enumitem}
\setlist{nosep}

\usepackage[normalem]{ulem}
\usepackage[numbers,sort]{natbib}

\usepackage{caption}
\usepackage{xspace}

\usepackage{pifont}

\begin{document}

\title{Browser Feature Usage on the Modern Web}

\author{
\numberofauthors{4}
\alignauthor  Peter Snyder\\
       \email{psnyde2@uic.edu}
\alignauthor Lara Ansari \\
\email{lansar2@uic.edu}
\alignauthor  Cynthia Taylor\\
       \email{cynthiat@uic.edu}
 \and
  \alignauthor  Chris Kanich\\
       \email{ckanich@uic.edu}
 \affaddr{Department of Computer Science}\\
 \affaddr{University of Illinois at Chicago}\\
\affaddr{Chicago, IL 60607}
}

\date{Draft: \today}

\maketitle

\newcommand{\CDF}{CDF\xspace}
\newcommand{\HTMLJS}{HTML and JavaScript\xspace}
\newcommand{\HTML}{HTML\xspace}
\newcommand{\BROWSERS}{commodity web browsers\xspace}
\newcommand{\JS}{JavaScript\xspace}
\newcommand{\XSS}{cross-site scripting\xspace}
\newcommand{\subsubsubsection}[1]{\paragraph*{\normalfont\normalsize\emph{#1}}}

\newcommand{\numfeatures}{1,392\xspace}
\newcommand{\numstandards}{74\xspace}
\newcommand{\numfirefoxes}{186\xspace}

\newcommand{\cvestotal}{470\xspace}
\newcommand{\cvesstandards}{111\xspace}

\newcommand{\FF}{Firefox\xspace}
\newcommand{\FFversion}{46.0.1\xspace}

\newcommand{\fixme}[1]{\textcolor{red}{#1}}
 \begin{abstract}
Modern web browsers are incredibly complex, with millions of lines of code and
over one thousand \JS functions and properties available to website authors.
This work investigates how these browser features are used on the modern, open
web.  We find that \JS features differ wildly in popularity, with over 50\% of
provided features never used in the Alexa 10k.

We also look at how popular ad and tracking blockers change the distribution of
features used by sites, and identify a set of approximately 10\% of features that
are disproportionately blocked (prevented from executing by these extensions at
least 90\% of the time they are used).  We additionally find that in the
presence of these blockers, over 83\% of available features are executed on
less than 1\% of the most popular 10,000 websites.

We additionally measure a variety of aspects of browser feature usage on the web,
including how complex sites have become in terms of feature usage, how the length
of time a browser feature has been in the browser relates to its usage on
the web, and how many security vulnerabilities have been associated
with related browser features.
 \end{abstract}

\section{Introduction}
The web is the world's largest open application platform.  While initially
developed for simple document delivery, it has grown to become the
most popular way of delivering applications to users.  Along with this growth
in popularity has been a growth in complexity, as the web has
picked up more and more capabilities over time.

This growth in complexity has been guided by both browser vendors and web
standards.  Many of these new web capabilities are provided through new
JavaScript APIs (referred to in this paper as \textbf{features}).  These
features are organized into collections of related 
features which are published as part of standards documents (in this paper, we
refer to these collections of APIs as \textbf{standards}).

To maximize compatibility between websites and web browsers,
browser vendors rarely remove features from browsers.  Browser vendors
 aim to provide website authors with new features without breaking
sites that rely on older browser features.  The result is an ever growing
set of features in the browser.

Many web browser features have been controversial and even actively opposed
by privacy and free software activists for imposing significant costs
on users, in the form of information leakage or loss of control.
The \emph{WebRTC}~\cite{webrtcw3c} standard has been
criticized for revealing users IP addresses~\cite{webrtcprivacy2015}, and
protestors have literally taken to the streets~\cite{emeprotests2016} to oppose the
\emph{Encrypted Media Extensions}~\cite{eme} standard. This standard aims to
give content owners much more control over how their content is experienced
within the browser. Such features could be used to prevent users from exerting
control over their browsing experience. 

Similarly, while the complexity (measured as the number of resources requested) is well
understood~\cite{butkiewicz2011understanding}, what is not understood is how
much of the functionality available in the browser gets used, by which
sites, how often, and for what purpose.  Several questions remain, including
whether recently introduced features are as popular as old features, whether
popular websites use different features than less popular sites, or how the use
of popular site altering extensions, like those that block advertisements and
online tracking, impact which browser features are used.

This paper answers those questions by examining the utilization of browser features on
the web.  By examining the \JS feature usage of the ten thousand most popular sites
on the web, we measured which browser features are frequently used by site authors,
and which browser features are rarely used on the web.  We find, for example, that
50\% of the JavaScript provided features in the web browser are never
used by the top ten thousand most popular websites.

Ad and tracking blocking extensions are a common way that users modify their
web browsing experience. We additionally measure the utilization of browser
features under these blockers to determine the change in browser feature usage
when users install these popular extensions. We find that installing
advertising and tracking blocking extensions not only unsurprisingly reduces
the amount of \JS users execute when browsing the web, but changes the kinds of
features their browsers execute; we identify a substantial set of browser
features (approximately 10\%) that are used by websites, but which ad and
tracking blockers prevent from executing more than ninety percent of the time.
Similarly, we find that over 83\% of features available in the browser are
executed on less than 1\% of websites in the presence of popular advertising
and tracking blocking extensions.

 \section{Background}
\label{sec:background}

In this section, we discuss the complexity of the modern web browser, along with the use of ad and tracking blockers.

\subsection{Modern Web Features}
\label{sec:instrumentation}

\begin{figure}[tb]
    \centering
\centerline{
    \includegraphics[width=\columnwidth]{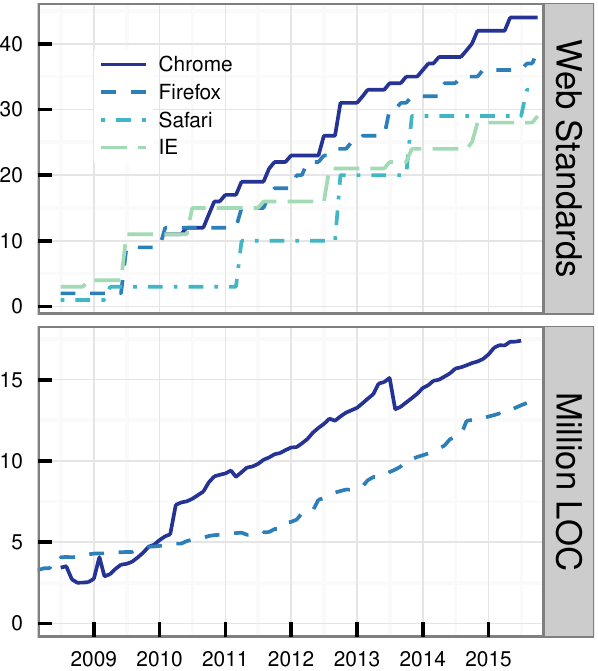}
}
    \caption{Feature families and lines of code in popular browsers over time.}
    \label{fig:1_browserfeatures}
    \vspace*{-0.1in}
\end{figure}

The complexity of modern web browsers has grown to encompass countless
potential use cases. While the core functionality embodied by the combination
of HTML, CSS, and JavaScript is largely stable, over the past few years,
many features have been added to provide for new use cases.
Figure~\ref{fig:1_browserfeatures} shows the number of standards available in
modern browsers, using data from W3C documents~\cite{w3cstandards} and
Can I Use~\cite{deveria2015caniuse}.
Figure~\ref{fig:1_browserfeatures} also shows the total number of lines of code
for Firefox and Chrome~\cite{openhubloc}. One relevant point of note in the figure
is that in mid 2013, Google moved to the Blink rendering engine, which entailed
removing at least 8.8 million lines of code from Chrome  related to the
formerly-used WebKit engine~\cite{blink-loc}.

Vendors are very wary of removing features from the browser, even if they are
used by a very small fraction of all websites~\cite{chromium-newfeatures,blinkdevmail}.
Because the web is evolving and competing with native applications, browser
vendors are incentivized to continue adding new features to the web browser and
not remove old features. This is exacerbated by browsers typically having a
unified code base across different types of computers including mobile devices,
browser-based computers such as Google Chromebooks, and traditional personal
computers. Browser vendors then expose unique hardware capabilities like
webcams, rotation sensors, vibration motors, or ambient light
sensors~\cite{webcamapi, rotationapi, vibrationapi,ambientlightapi} directly
through JavaScript, regardless of whether the executing device has such a
capability.  Furthermore, as new features are added, the current best practice
is to roll them out directly to web developers as time limited experiments, and
allow them to move directly from experimental features to standard features,
available in all browsers that adhere to the HTML living standard.~\cite{russel2015doing}.

Individual websites are also quite complex.  Butkiewicz et al. surveyed 2000
random websites and found that loading the base page for a URL required
fetching a median of 40 objects, and that 50\% of websites fetched at least 6
Javascript objects~\cite{butkiewicz2011understanding}.

\vfill\eject
\subsection{Ads and Tracking Blocking}

Researchers have previously investigated how people use ad blockers.
Pujol et al. measured AdBlock usage in the wild, discovering that while a
significant fraction of very active web users use AdBlock, most users
primarily use its ad blocking, and not its privacy preserving,
features~\cite{pujolannoyed}.

User tracking is a more insidious aspect of the modern web. Recent work by Radler
found that users were much less aware of cross-website tracking than they were
about collection of data by single sites such as Facebook and Google, and that
users who were aware of it had greater concerns about unwanted access to
private information than those who weren't aware~\cite{rader2014awareness}.
Tracking users' web browsing activity across websites is largely unregulated,
and a complex network of mechanisms and businesses have sprung up to provide a
variety of services in this space~\cite{falahrastegar2014anatomy}.
Krishnamurthy and Willis found that aggregation of user-related data is both
growing and becoming more concentrated, i.e. being conducted by a smaller number
of companies~\cite{krishnamurthy2009privacy}.

Traditionally, tracking was done via client-side cookies, giving users a measure
of control over how much they are tracked (i.e. they can always delete cookies).
However, a wide variety of non-cookie tracking measures have been developed that
take this control away from users, and these are what tracking blockers have
been designed to prevent. These include browser
fingerprinting~\cite{eckersley2010unique}, JavaScript
fingerprinting~\cite{mowery2011fingerprinting,mulazzani2013fast}, Canvas
fingerprinting~\cite{mowery2012pixel}, clock skew
fingerprinting~\cite{kohno2005remote}, history
sniffing~\cite{jang2010empirical}, cross origin timing attacks~\cite{van2015clock},
evercookies~\cite{evercookies}, and Flash cookie
respawning~\cite{soltani2010flash,ayenson2011flash}.  A variety of these
tracking behaviors have been observed in widespread use in the
wild~\cite{soltani2010flash,ayenson2011flash,acar2014web,
nikiforakis2013cookieless,mcdonald2011survey,olejnik2014selling,sorensen2013zombie}.

Especially relevant to our work is the use of JavaScript APIs for tracking.
While some APIs, such as Beacon~\cite{beaconapi}, are designed specifically for
tracking, other APIs were designed to support various other functionality and
co-opted into behaving as trackers~\cite{mowery2012pixel,fingerprintjs2}.
Balebako et al. evaluated tools which purport to prevent tracking and found that
blocking add-ons were effective~\cite{balebako2012measuring}.
 \section{Data sources}
\label{sec:data-sources}
This work draws on several existing sets of data.  This section proceeds by
detailing how we determined which websites are more popular and how often
they are visited, how we determined the current \JS-exposed feature set
of a modern web browser, what web standard those features belong to and when
they were introduced, how we determined the known vulnerabilities in the web
browser (and which browser feature standard the vulnerability was associated
with), and which browser plugins we used as representative of common browser
modifications.

\subsection{Alexa Website Rankings}
\label{sec:website-popularity-rankings}
The Alexa rankings are a well known ordering of websites ranked by traffic.
Typically, research which uses Alexa relies on their ranked list of the
worldwide top one million sites. However, Alexa exposes more data about these
sites through their API. In addition to a global ranking of each of these sites,
there are also local rankings at country granularity, breakdowns of which
subsites (by fully qualified domain name) are most popular, and a breakdown by
page load and by unique visitor of how many monthly visitors each site gets.

We use the Alexa rankings to determine the 10,000 most popular sites,
which collectively represent approximately one third of all web visits.

\subsection{Web API Features}
\label{sec:method-web-standards}
We define a \textbf{feature} as a browser capability that is accessible through a
\JS function call or property setting.

We determined the set of \JS-exposed browser features by
reviewing the WebIDL definitions included in the Firefox version 46.0.1 source
code. WebIDL is a language that defines the \JS features web browsers
provide to web authors.  In the case of Firefox, these WebIDL files are
included as text documents in the browser source.

In the common case, Firefox's WebIDL files define a mapping between a
\JS accessible method or property and the C++ code that implements
the underlying functionality in the browser\footnote{In addition to mapping
\JS to C++ methods and structures, WebIDL can also define \JS
to \JS methods, as well as intermediate structures that are not
exposed to the browser.  In practice though, the primary role of WebIDL
in Firefox is to define a mapping between \JS API endpoints and
the underlying implementations, generally in C++.}. We examined each of the 757
WebIDL files in the Firefox and extracted \numfeatures relevant methods and
properties implemented in the browser.

\subsection{Web API Standards}
Web standards are documents defining functionality that web browser vendors
should implement.  They are generally written and formalized by organizations
like the W3C, though occasionally standards organizations delegate
responsibility for writing standards to third parties, such as the Khronos
group (who maintains the current WebGL standard).

Web API standards are a collection of one or more Web API features,
generally designed to be used together.  For example, the
\emph{WebAudio API}~\cite{webaudio2013standard} standard defines 52 \JS APIs
that together allow page authors to do programmatic sound synthesis.

While there are web standards that cover many aspects of the web (such
as parsing rules, what tags and attributes can be used in HTML documents,
etc.) this work focuses only on web standards that define \JS exposed
functionality.

Web API standards documents enumerate the WebIDL endpoints that are part of
each standard.
We identified \numstandards standards implemented in \FF.  We associated each
of the \numfeatures features we identified to one of these standards.
We also found 65 API endpoints implemented in Firefox that are not
found in any web standard document, which we associated with a catch-all
\emph{Non-Standard} categorization.

In the case of some extremely large standards we identify sub-standards,
which define a subset of related features intended to be used together.
For example, we treat the subsections of the HTML standard that define the
basic Canvas API, or the WebSockets API, as their own standards.

Because these sub-standards have their own coherent purpose,
it is meaningful to discuss them independently of their parent standards. Furthermore,
many have historically been implemented in browsers independent
of the parent standard (i.e. browser vendors added support for
``websockets'' long before they implement support for the current full ``HTML5''
standard.

Some features appear in multiple web standards.  For example, the
\texttt{Node.prototype.insertBefore} feature appears in the
\emph{Document Object Model (DOM) Level 1 Specification}~\cite{dom1w3c},
\emph{Document Object Model (DOM) Level 2 Core Specification}~\cite{dom2corew3c}
and \emph{Document Object Model (DOM) Level 3 Core Specification}~\cite{dom3corew3c}
standards.  In such cases, we attribute the feature to the earliest published
standard.

\subsection{Historical Firefox Builds}
We determined when features were implemented in Firefox by
examining the \numfirefoxes versions of \FF that have been released since 2004
and testing when each of the \numfeatures features we identified in the current version
\FF first appeared.  We treat the release date of the earliest verison of
\FF that a feature appears in as the feature's ``implementation date''.

Most standards do no have a single implementation date, since it could take
months or years for a standard to be fully implemented in \FF.  We therefore
treat the introduction of a standard's currently most popular feature as the
standard's implementation date. For ties (especially relevant when no feature
in a standard is used), we default to the earliest feature available.

\subsection{CVEs}
\label{sec:data-cves}
We collected information about browser vulnerabilities and security bugs by
finding all Common Vulnerabilities and Exposures
(CVEs)~\cite{cvedatabase} (security-relevant bugs discovered in
software) related to the Firefox web browser that had been documented in the
last three years.

The CVE database lists \cvestotal issues from the last three years that mention
Firefox.  On manual inspection we found that 14 of these were not actually
issues in Firefox, but issues in other web-related software where Firefox
was used to demonstrate the vulnerability.

Of the remaining 456 CVEs, we were able to manually associate \cvesstandards
CVEs with a specific web standard.  For example, CVE-2013-0763~\cite{firefoxWebglCVE}
describes a potential remote execution vulnerability introduced in Firefox's
implementation of the \emph{WebGL}~\cite{webgl2015standard} standard, and
CVE-2014-1577~\cite{firefoxWebAudioCVE} documents a potential information-disclosing bug related to Firefox's implementation of
the \emph{Web Audio API} standard.

\subsection{Blocking Extensions}
\label{sec:data-extensions}
Finally, this work pulls from commerical and crowd-sourced browser extensions, which are popularly used to modify the
browser environment in a way that the user prefers.

This work relies on two such browser extensions, Ghostery and AdBlock Plus.  Ghostery is a
browser extension that allows users to increase their privacy online
by modifying their browser to not load resources or set cookies associated with
cross-domain passive tracking, as determined by the extension's maintainer, Ghostery, Inc..

This work also uses the AdBlock Plus browser extension, which modifies
the browser to both not load resources the extension associates with
advertising, as well as to hide elements in the page that are advertising related.
This extension draws from a crowdsourced list of rules and URLs that the
extension uses to determine whether a resource is advertising-related.

 \section{Methodology}
\label{sec:methodology}

To understand browser feature usage on the web, we conducted a survey of the
Alexa 10k, visiting each site ten times and recording which browser features
were utilized.  We visited each site five times with an unmodified browsing
environment, and five times with popular tracking-blocking and
advertising-blocking extensions installed.  This section proceeds by describing
the goals of this survey, followed by how we instrumented the web browser to
determine which features are used on a given site, and then concludes with how
we used our instrumented browser to measure feature usage on the web.

\subsection{Goals}
The goal of our automated survey is to take a cross section of the web as it is
commonly experienced by users, and to determine which browser features
are used in those websites.  This requires us to take a
broad-yet-representative sample of the web, and to exhaustively determine the
functionalities most commonly used on those sites.

To do so, we built a browser extension to measure which features are used when a user interacts with a website.
We then chose a representative sample of the web to visit.  Finally, we developed a method for interacting with these sites in an
automated fashion to elicit the same functionality that a human web
user would experience.  Each of these steps is described in further detail
in the proceeding subsections.

This automated approach only attempts to measure the ``open web'', or the
subset of webpage functionality that a user encounters \emph{without} logging
into a website.  Users may encounter different types of functionality when
interacting with websites that they have created accounts for and established
relationships with, but such measurements are beyond the scope of this paper.

\subsection{Measuring Extension}
\label{sec:measureextension}

We instrumented a recent version of the Firefox web
browser (version \FFversion) with a custom browser extension which
records each time a \JS feature has been used on a visited page.  Our extension
injects \JS into each page after the browser has created the DOM for that page,
and before any of the page's content has been loaded. By injecting this \JS
into the beginning of the \texttt{<head>} element, we
 can modify those methods and properties comprising the DOM before it becomes
available to the \JS of the requested page.

\begin{figure}[tb]
    \centering
\centerline{
    \includegraphics[width=\columnwidth]{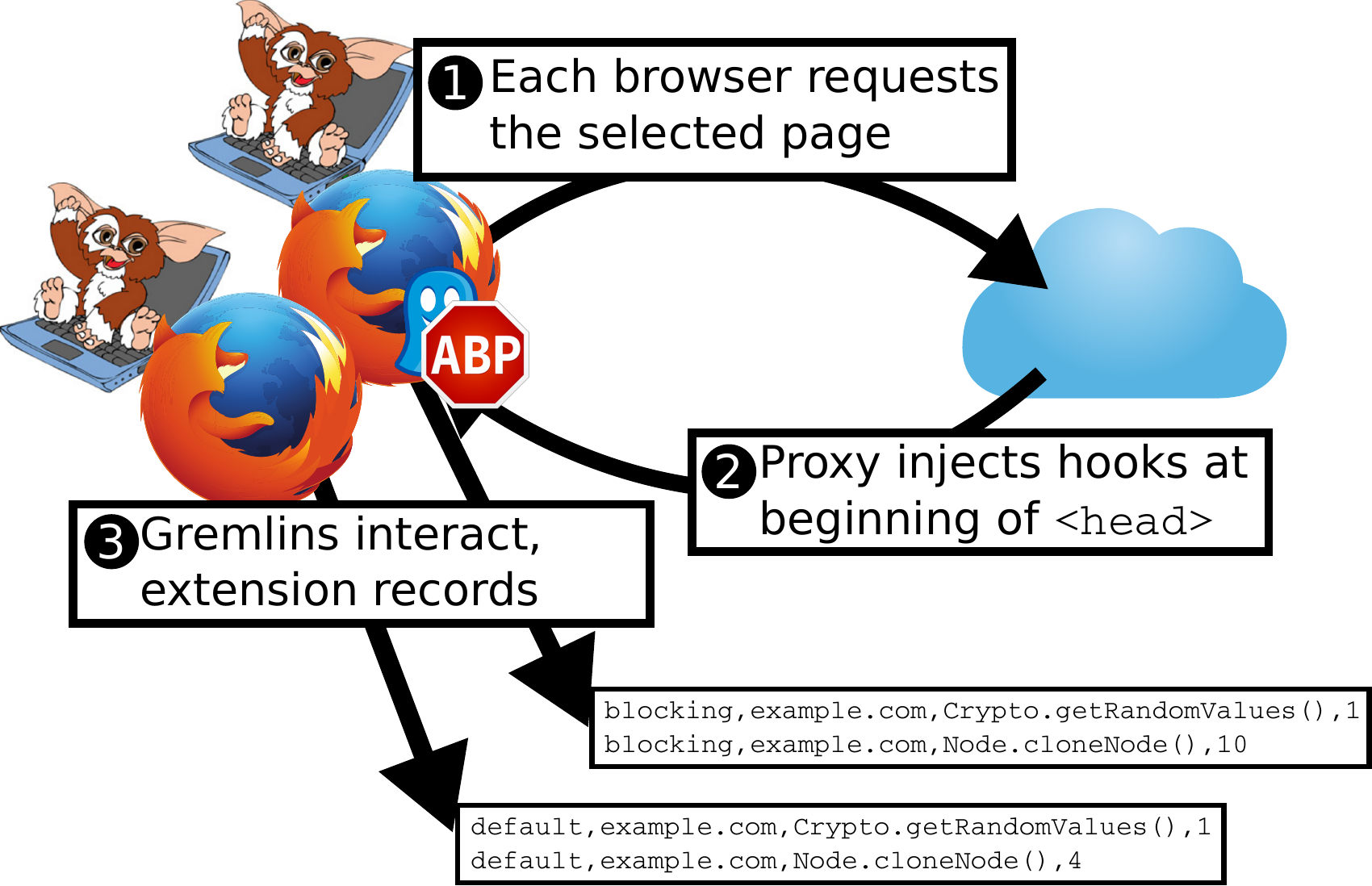}
}
    \caption{One iteration of the feature invocation measurement process.}
    \label{fig:gremlins}
    \vspace*{-0.1in}
\end{figure}

The \JS that the extension injects into each requested page modifies
the DOM to count each instance an instrumented method is called or that an
instrumented property is written.  How the extension measures these method calls
and property writes is detailed in the following two subsections.
Figure~\ref{fig:gremlins} presents a representative diagram of the crawling process.

\subsubsection{Measuring Method Calls}
The browser extension counts method invocations by overwriting each method on
the defining element's prototype.  This approach allows us to shim in our own
logging functionality for each method call, and then call the original method
to preserve the original functionality.  This replaces each reference to the
DOM's methods with the extension's instrumented version.

We also take advantage of closures in \JS to ensure that web
pages are not able to bypass the instrumented versions of each method by
looking up--or otherwise directly accessing--the original versions of each
DOM method.

\subsubsection{Measuring Property Writes}
Properties were more difficult to record.  \JS provides no way to
intercept whether a property has been set or read on a client script-created object,
or on an object created after the instrumenting code has finished executing.
However, through the use of the non-standard
\texttt{Object.watch()}\cite{mozillaobjectwatch} method available in
Firefox, we were able to capture property-setting events on
one of the singleton objects in the browser (e.g. \texttt{window},
\texttt{window.document}, \texttt{window.navigator}).  Using this
\texttt{Object.watch()} method allowed the extension to capture and count
all writes to properties on singleton objects in the DOM.

\subsubsection{Other Browser Features}
Web standards define other features in the browser too, such as events
and CSS layout rules, selectors, and instructions.  However, our extension-based
approach did not allow us to measure the use of these features, and so counts
of their use are not included in this work.

In the case of standard defined browser events, the extension could have
captured some event registrations by combination of watching which
events were registered to \texttt{addEventListener} method calls,
and watching for property-sets to singleton objects.  However, we would not
have been able to capture event registrations using the legacy \texttt{DOM0}
method of event registration (e.g. assigning a function to an object's
\texttt{onclick} property to handle click events) on non-singleton objects.
Since we would only have been able to see a subset of event registrations,
we decided to omit events from this work.

Similarly, this work does not consider non-\JS exposed functionality
defined in the browser, such as CSS selectors and rules.  While interesting,
this work focuses solely on functionality that the browser allows \JS
authors to access.

\subsection{Eliciting Site Functionality}
\label{sec:eliciting-site-functionality}

Using our feature-detecting browser extension, we were able to measure which browser features are used on
the 10k most popular websites.  The following subsections describe how we simulated human interaction
with web pages to capture the features that would be executed when users
visited these sites, first with the browser in its default state, and again
with the browser modified with popular advertisting-and-tracking blocking
extensions.

\subsubsection{Default Case}
\label{sec:default-case-measurements}

To understand which features are necessary for a site's execution, we perform
dynamic analysis on live running pages in \FF while using the measuring
extension described in Section~\ref{sec:measureextension}.  The goal is to
exercise as much of the functionality built into the page as possible.
While some \JS features of a site are automatically activated on the home page,
like advertisements and analytics,  many features will only be used as a result
of user interaction either within the page or by navigating to different areas
of the site. Here we explain our strategy for crawling and interacting with sites.

In order to trigger as many browser features as possible on a website, we
used a common site testing methodology called ``monkey testing''.  Monkey
testing refers to the strategy of instrumenting a page to click, touch, scroll,
and enter text on random elements or locations on the page.  To accomplish this, we use a
modified version of gremlins.js~\cite{zaninotto2016gremlins}, a library built
for monkey testing front-end website interfaces.

We started our measurement by visiting the home page of each site and allowing
the monkey testing to run for 30 seconds.  Because the randomness of monkey
testing could cause navigation to other domains, we intercepted and prevented
any interactions which might navigate to a different page.  For
navigations that would have been to the local domain, we noted which URLs the
browser would have visited in the absence of the interception.

We then proceeded in a breadth first search of the site's hierarchy using the
URLs that would have been visited by the actions of the monkey
testing.  We then selected 3 of these URLs that were on the same domain (or
related domain, as determined by the Alexa data), and visited each, repeating
the same 30 second monkey testing procedure and recording all used features.
From each of these 3 sites, we then visited three more pages for 30 seconds,
which resulted in a total of 13 pages interacted with for a total of 390 seconds
per site.

If more than three links were clicked during any stage of the monkey testing
process, we selected which URLs to visit by  giving preference to URLs where
the directory structure of the URL had not been previously seen. In contrast to
traditional interface fuzzing techniques which have as a goal finding unintended
or malicious functionality~\cite{amalfitano2012using,liu2014decaf}, we were
interested in finding all functionalities that users will commonly interact
with.  By selecting URLs with different path-segments, we try to visit as many
different types of pages on the site as possible, with the goal of
capturing all of the functionality on the site that a user would encounter.
These properties of our strategy are evaluated in
Section~\ref{sec:validation}.

\subsubsection{Blocking Case}
\label{sec:blocking-case-measurements}
In addition to the default case measurements described in
Section~\ref{sec:default-case-measurements}, we also re-ran the same measurements
against the Alexa 10k with an ad blocker (AdBlockPlus) and a tracking-blocker
(Ghostery) to generate a second, `blocking', set of measurements. We include
these blocking measurements as being representative of the types of modifications
users make to customize their browsing experience. While this version of a site
no longer represents its author's intended representation (and may in fact
break the site), the popularity of these content blocking extensions lends
credence to the blocking case being a valid alternative method of experiencing
a given website.

\subsubsection{Automated Crawl}
\label{sec:automated-crawl}

\begin{table}[h]
  \centering
  \begin{tabular}{ l r }
    \toprule
      Domains measured       &  9,733 \\
      Total website interaction time      &  480 days \\
      Web pages visited                   &  2,240,484    \\
      Feature invocations recorded        &  21,511,926,733 \\
    \bottomrule
  \end{tabular}
  \caption{Amount of data gathered regarding \JS feature usage on
the Alexa 10k.
``Total website interaction time'' is an estimate based on
the number of pages visited and 30 seconds of page interaction per visit.}
  \label{fig:results-vanity-stats}
\end{table}

For each site in the Alexa 10k, we repeated the above procedure ten times
to ensure we measure all features used on the page, five times in the default
case, and 5 time in the blocking case.  We present findings for why
5 times is sufficient to induce all types of site functionality in
Section~\ref{sec:validation}.  Table~\ref{fig:results-vanity-stats} presents
some high level figures of this automated crawl.  For 267 domains,
were unable to measure feature usage for a variety of reasons, including
non-responsive domains and sites that contained syntax errors in
their \JS code that prevented execution.
 \section{Results}
\label{sec:results}

In this section we discuss our findings, including the popularity distribution
of \JS features used on the web with and without blocking, feature popularity's
relation to feature age, which features are disproportionately blocked, and
which features are associated with security vulnerabilities.

\subsection{Definitions}
This work uses the term \textbf{feature popularity} to denote the percentage of
sites that use a given feature at least once during automated interaction with
the site.  A feature that is used on every site has a popularity of 1, and
a feature that is never seen has a popularity of 0.

Similarly, we use the term \textbf{standard popularity} to denote the
percentage of sites that use at least one feature from the standard at least
once during the site's execution.

Finally, we use the term \textbf{block rate} to denote how
frequently a browser feature would have been used if not for the presence of an
advertisement- or tracking-blocking extension. In other words, browser features
that are used much less frequently on the web when a user has AdBlock Plus or
Ghostery installed have high block rates, while features that are still used on
roughly the same number of websites in the presence of blocking extensions have
low block rate.

\subsection{Standard Popularity Distribution}

\begin{figure}[th]
    \centering
    \centerline{
        \includegraphics{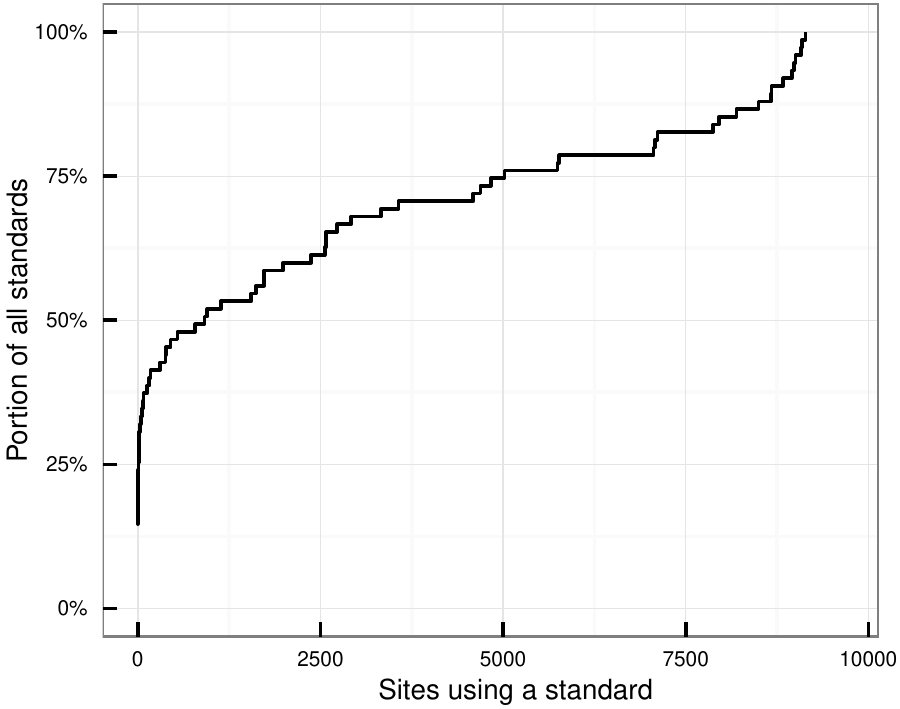}
    }
    \caption{Cumulative distribution of standard popularity within the Alexa 10k.}
    \label{fig:popcdf}
    \vspace*{-0.1in}
\end{figure}

Figure~\ref{fig:popcdf} displays the cumulative distribution of standard
popularity. Some standards are extremely popular, and others are extremely
unpopular: six standards are used on over 90\% of all websites
measured, and a full 28 of the 75 standards measured were used on 1\% or fewer
sites, with eleven not used at all. Standard popularity is not feast or famine
however, as standards see several different popularity levels between those two
extremes.

\subsection{Feature Popularity}
We find that browser standards are not equally used on the web.  Some features
are extremely popular, such as the \texttt{Document.prototype.createElement}
method, which allows for client-side modification of webpages and is used on 9,079--or over 90\%--of
pages in the Alexa 10k.

Other browser features are never used.  For example, 689 features, or almost
50\% of the \numfeatures implemented in the browser, are never used once in the
10k most popular sites.  A further 416 features are used on less than 1\% of the
10k most popular websites.  Put together, this means that over 79\% of the
features available in the browser are used by less than 1\% of the web.

We also find that browser features do not have equal block rates; some features
are blocked by advertisement and tracking blocking extensions far more often
than others.  10\% of browser features are prevented from executing
over 90\% of the time when browsing with common blocking extensions.   We
also find that 1,159 features, or over 83\% of features available in the browser,
are executed on less than 1\% of websites in the presence of popular advertising
and tracking blocking extensions.

\subsection{Standard Popularity in the Alexa 10k}
\label{sec:results-feature-popularity}

\begin{figure*}[htb]
    \centering
    \centerline{
        \includegraphics{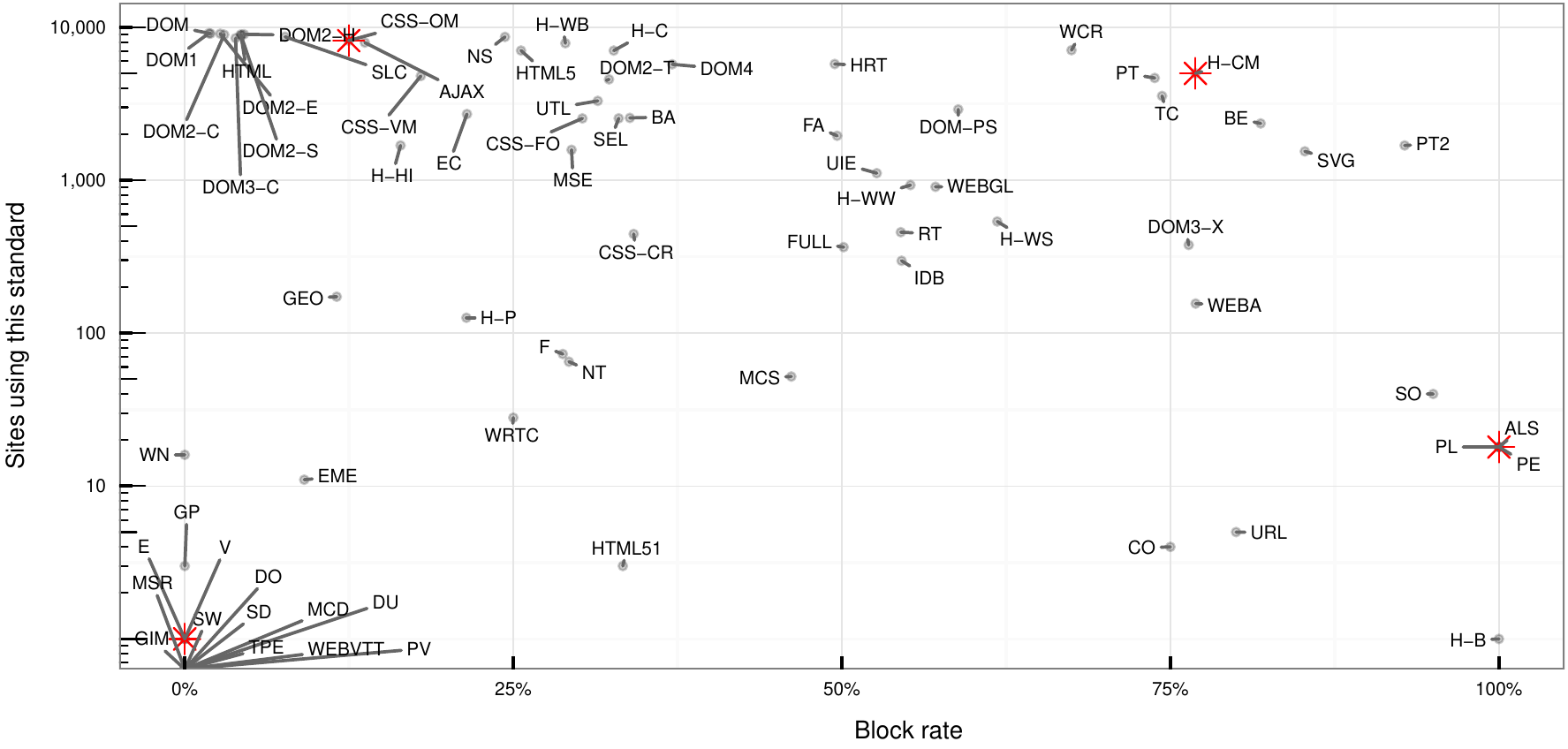}
    }
    \caption{Popularity of standards versus their block rate, on a log scale.}
    \label{fig:megagraph}
    \vspace*{-0.1in}
\end{figure*}

Standards, sets of related browser features collected and standardized together,
are not equally popular on the web.  Figure \ref{fig:megagraph}
depicts the relationship between a standards's popularity (represented by
the number of sites the standard was used on, log scale) and
its block rate.  Since a standard's popularity is the number of sites where
a feature in a standard appears at least once, the popularity of the standard
will be equal to at least the popularity of the most popular feature in the standard.

Each quadrant of the graph tells a different story about the popularity and
desirability of a standard on the web.

\subsubsubsection{Popular, Unblocked Standards}
The upper-left quadrant contains the
standards that occur very frequently on the web, and are rarely blocked by
advertising and tracking blocking extensions.

One example of this, point \textbf{CSS-OM}, depicts the
\emph{CSS Object Model}~\cite{cssomw3c} standard, which allows \JS code
to introspect, modify and add to the styling rules in the document.  It is
positioned near the top of the graph because 8,193 sites used a feature
from the standard at least once during measurement.  The standard is positioned
to the left of the graph because the standard has a low block rate (12.6\%),
meaning that the addition of blocking extensions had relatively little effect
on how frequently a site used any feature from the standard.

\subsubsubsection{Popular, Blocked Standards}
The upper-right quadrant of the graph shows standards that are used by a
large percentage of sites on the web, but which blocking extensions frequently
prevent from executing.

A representative example of such a standard is the \emph{HTML: Channel Messaging}
~\cite{htmlcmw3c} standard, represented by point \textbf{H-CM}.  This standard
defines \JS methods allowing embedded documents (\texttt{iframes}) and windows
to communicate with their parent document.  This functionality is often used
by embedded-content and popup windows to communicate with the hosting page,
often in the context of advertising.  This standard is used on over half of
all sites by default, but is prevented from being executed over 77\% of the
time.

\subsubsubsection{Unpopular, Blocked Standards}
The lower-right quadrant of the graph shows standards that are rarely used by
websites, and that are almost always prevented from executing by blocking extensions.

Point \textbf{ALS} shows the \emph{Ambient Light Events}
standard~\cite{ambientlightapi}, which defines events and methods allowing
a website to react to changes to the level of light the computer, laptop
or mobile phone is exposed to.  The standard is rarely used on the web (14 out of
10k sites), but is prevented from being executed 100\% of the time by blocking
extensions.

\subsubsubsection{Unpopular, Unblocked Standards}
The lower-left quadrant of the graph shows standards that were rarely seen
in our study, but which were rarely prevented from executing.  Point \textbf{E}
shows the \emph{Encodings}~\cite{encodingw3c} standard.  This
standard allows \JS code to read and convert text between different text
encodings, such as reading text from a document encoded in \emph{GBK}
and inserting it into a website encoded in \emph{UTF-8}.

The \emph{Encodings}~\cite{encodingw3c} standard is rarely used on the web, with only 1 of the Alexa 10k sites
attempting to use it.  However, the addition of an advertising or
tracking blocking extension had no affect on the number of times the standard
was used; this sole site still used the \emph{Encodings} standard although
AdBlock Plus and Ghostery were installed.

\subsection{Standard Popularity by Site Popularity}
\label{sec:results-feat-pop-by-site-pop}

\begin{figure}[h]
  \centering
  \includegraphics[width=\columnwidth]{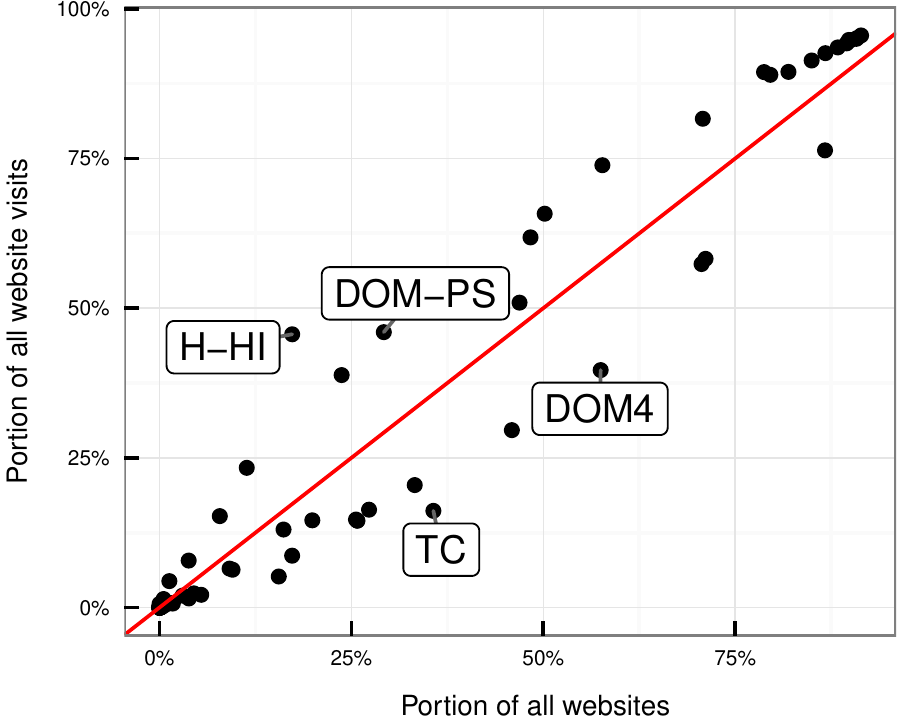}
  \caption{Comparison of percentage of sites using a standard versus percentage of web traffic using a standard.}
  \label{fig:feature-pop-by-site-pop}
\end{figure}

The results described in this paper treat all sites in the Alexa 10k equally,
so that if the most popular and least popular sites use the same standard,
both uses of that standard are given equal consideration.  In this section
we examine the accuracy of this assumption by measuring the difference between
the number of sites using a standard, and the number of website visits using a standard.
In other words, we weigh a standard's use based on the popularity of the site using it.

Figure \ref{fig:feature-pop-by-site-pop} shows the results of this comparison.
The x-axis shows the percentage of sites examined that use at least one
feature from a standard, and the y-axis shows the estimated percentage of
site views on the web that use this standard.  Standards above the
\texttt{x=y} line are more popular on frequently visited
sites, such that the percentage of page views using the standard is greater
than the percentage of sites using the standard.

Generally, the graph shows that standard usage is not equally distributed, and
that some standards are more popular with more frequently visited sites.
However, the general trend appears to be for standards to
cluster around the \texttt{x=y} line, indicating that while there are some
differences in standard usage between popular and less popular sites, they do
not affect our general analysis of standard usage on the web.

Therefore, for the sake of brevity and simplicity, all other measures in this
paper treat standard usage on all domains as equal, and do not factor a site's
popularity into the measurement.

\vfill\eject
\begin{figure}[h]
  \centering
  \includegraphics[width=\columnwidth]{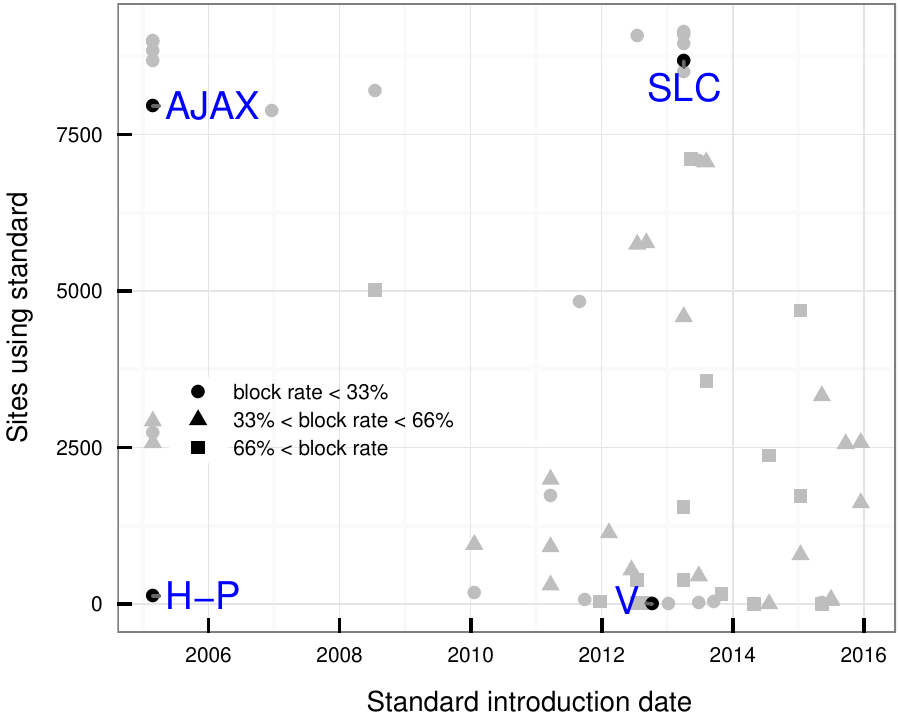}
  \caption{Comparison of a standard's availability date, and its popularity.}
  \label{fig:date_popularity}
\end{figure}

\subsection{Standard Popularity by Introduction Date}

We were also able to measure the relationship between when a standard became
available in the browser, its popularity, and how frequently its execution is
prevented by popular blocking extensions.  Again, to simplify
the presentation, the graph illustrates the popularity, introduction date, and block
rate of standards; not of the individual features themselves that comprise these standards.

As the graph shows, there is no simple relationship between when a standard
was added to the browser, how frequently the standard is used on the web
today, and how frequently the standard is blocked by common blocking
extensions.  However, as Figure~\ref{fig:date_popularity} indicates, some standards
have become extremely popular over time, while others, both recent and old,
have languished in disuse. Further, it appears that some standards have been
introduced extremely recently but have nevertheless been readily adopted by web authors.

\ \\*\emph{Old, Popular Standards}. For example, point \textbf{AJAX} depicts the \emph{XMLHttpRequest}~\cite{ajaxwhatwg},
or \emph{AJAX} standard, used to to send information to a server
without refetching the entire document.  This standard has been available in
the browser for almost as long as \FF has been released (since 2004), and is
also extremely popular; this standard's most popular feature, \texttt{XMLHttpRequest.prototype.open},
is used on 7,955 sites in the Alexa 10k.  Standards in this portion of the graph have been in the browser
for a long time, and appear on a large fraction of sites.  This cluster of
standards also have generally low block rates of less than 50\%, which are considered low in this study.

\ \\*\emph{Old, Unpopular Standards}. Other standards, despite existing in the browser nearly since \FF's inception,
are much less popular on the current web.  Point \textbf{H-P} shows the
\emph{HTML: Plugins}~\cite{htmlpluginsw3c} standard, which is a subsection of the
larger \emph{HTML} standard that allows document authors to detect the
names and capabilities of plugins installed in the browser (such as
Flash, Shockwave, Silverlight, etc.).  The most popular features of this standard
have been available in \FF since 2005.  However, the most popular feature in this
standard, namely, \texttt{PluginArray.prototype.refresh}, which checks for
changes in browser plugins, is used on less than 1\% of current websites (90 sites).

\ \\*\emph{New, Popular Standards}. Point \textbf{SEL} depicts the \emph{Selectors API Level 1}~\cite{selectors1w3c}
standard, which provides site authors with a simpler interface for querying
elements in a document.  Despite being a relatively recent addition to the
browser (the standard was added in 2013), the most popular feature in the
standard--the \texttt{Document.prototype.querySelectorAll} feature--is used on
over 80\% of websites.  This standard, along with the other standards
the \emph{Selectors API Level 1} standard is clustered with in the graph,
have low block rates.

\ \\*\emph{New, Unpopular Standards}. Point \textbf{V} shows the \emph{Vibration}~\cite{vibrationapi}
standard, which allows site authors to trigger a vibration in the user's device
on platforms that support it.  Despite this standard having been available in \FF
longer than the previously mentioned \emph{Selectors API Level 1} standard,
the \emph{Vibration} standard is significantly less popular on the web.  The
sole method in the standard, \texttt{Navigator.prototype.vibrate}, is only used
once in the Alexa 10k.

\subsection{Standard Blocking}
\label{sec:results-feature-blocking}
Many users alter their browsing environment when visiting websites.  They may
do so for a variety of reasons, including wishing to limit advertising displayed
on the pages they read, reducing their exposure to malware distributed through
advertising networks, and increasing their privacy by reducing the amount
of tracking they experience online.  These browser modifications are typically
made by installing browser extensions.

We measured the effect that installing two common browser
extensions, AdBlock Plus and Ghostery, on the type and number of
features that are executed when visiting websites.

\vfill\eject
\subsubsection{Standard Blocking Frequency}
As discussed in \ref{sec:results-feature-popularity}, browser standards are
not equally prevented from executing by installing blocking extensions.
As Figure \ref{fig:megagraph} shows, some standards are greatly impacted by
installing these advertising and tracking blocking extensions, while others
are generally not impacted at all.

For example, the \emph{Beacon}~\cite{beaconapi} standard, which allows websites
to trigger functionality when a user leaves a page, has a 83.6\% reduction
in usage when browsing with blocking extensions.  Similarly, the
\emph{SVG} standard, which includes functionality that allows for fingerprinting
users through font enumeration\footnote{The \texttt{SVGTextContentElement.prototype.\\
getComputedTextLength}
method}, sees a similar 86.8\% reduction in site usage when browsing with
blocking extensions.

Other browser standards, such as the core \emph{DOM} standards,
see very little reduction in page usage in the presence of blocking extensions.

\subsubsection{Standard Blocking Purpose}
\begin{figure}
  \centering
  \includegraphics[width=\columnwidth]{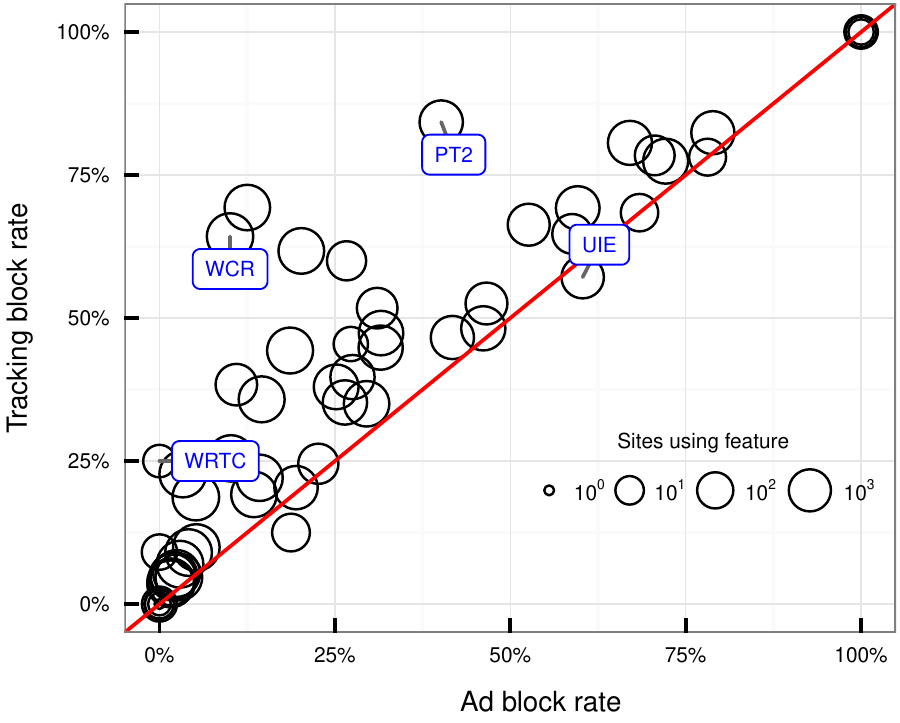}
  \caption{Comparison of block rates of standards using advertising vs. tracking blocking extensions.}
  \label{fig:feature-blocking-source}
\end{figure}

In addition to measuring which standards were blocked by extensions, we were
also able to distinguish which extensions did the blocking.
Figure~\ref{fig:feature-blocking-source} plots the block rate of standards
when sites were visited with only an advertising blocking extension
installed (x-axis), versus the block rate of standards when sites were
visited with only a tacking blocking extension installed (y-axis).

Values along the \texttt{X=Y} line in the graph are standards that were blocked
equally with both types of extensions installed, with points closer to the
upper-right corner being blocked more often, and points closer to the
lower-left corner being blocked less often.

Points to the upper-left of the graph depict standards that were blocked
more frequently by the tracking-blocking extension than the
advertising-blocking extension, while points to the lower-right of the graph
shows standards that were blocked more frequently by the advertising-blocking
extension.

As the graph shows, some standards, such as \emph{WebRTC}~\cite{webrtcw3c}
(which is associated with attacks revealing the user's IP address),
\emph{WebCrypto API}~\cite{webcryptow3c} (which is used by some analytics libraries
to generate identifying nonces), and
\emph{Performance Timeline Level 2}~\cite{perftimingw3c} (which is used to generate
high resolution time stamps) are blocked by tracking-blocking extensions
more often than they are blocked by advertisement blocking extensions.

The opposite is true, to a lesser extent, for the \emph{UI Events
Specification}~\cite{uievents3c} standard, which specifies new ways that
sites can respond to user interactions.

\subsection{Standards and Browser Vulnerabilities}
\begin{table*}[hp]
  \centering
  \rowcolors{2}{gray!25}{white}
  \begin{tabular}{ l l r r r r }
    \toprule
      Standard Name &
      Abbreviation &
      \multicolumn{1}{c}{\# Features} &
      \multicolumn{1}{c}{\# Sites} &
      \multicolumn{1}{c}{Block Rate} &
      \multicolumn{1}{c}{\# CVEs} \\
    \midrule
      HTML: Canvas                                                  & H-C    & 54    & 7,061   & 33.1\%   & 15 \\
      Scalable Vector Graphics 1.1 (2\textsuperscript{nd} Edition)  & SVG    & 138	  & 1,554   & 86.8\%	 & 14 \\
      WebGL	                                                        & WEBGL  & 136   & 913	    & 60.7\%	 & 13 \\
      HTML: Web Workers	                                            & H-WW   & 2	    & 952	    & 59.9\%	 & 11 \\
      HTML 5		                                                    & HTML5  & 69	  & 7,077	  & 26.2\%	 & 10 \\
      Web Audio API		                                              & WEBA   & 52    & 157	    & 81.1\%   & 10 \\
      WebRTC 1.0                                                    & WRTC   & 28    & 30	    & 29.2\%   & 8 \\
      XMLHttpRequest                                                & AJAX   & 13    & 7,957   & 13.9\%   & 8 \\
      DOM                                                           & DOM    & 36    & 9,088   & 2.0\%    & 4 \\
      Indexed Database API                                          & IDB    & 48    & 302     & 56.3\%   & 3 \\
      Beacon                                                        & BE     & 1     & 2,373   & 83.6\%   & 2 \\
      Media Capture and Streams                                     & MCS    & 4     & 54      & 49.0\%   & 2 \\
      Web Cryptography API                                          & WCR    & 14    & 7,113   & 67.8\%   & 2 \\
      CSSOM View Module                                             & CSS-VM & 28    & 4,833   & 19.0\%   & 1 \\
      Fetch                                                         & F      & 21    & 77      & 33.3\%   & 1 \\
      Gamepad                                                       & GP     & 1     & 3       & 0.0\%    & 1 \\
      High Resolution Time, Level 2                                 & HRT    & 1     & 5,769   & 50.2\%   & 1 \\
      HTML: Web Sockets                                             & H-WS   & 2     & 544     & 64.6\%   & 1 \\
      HTML: Plugins                                                 & H-P    & 10    & 129     & 29.3\%   & 1 \\
      Web Notifications                                             & WN     & 5     & 16      & 0.0\%    & 1 \\
      Resource Timing                                               & RT     & 3     & 786     & 57.5\%   & 1 \\
      Vibration API                                                 & V      & 1     & 1       & 0.0\%    & 1 \\
      Battery Status API                                            & BA     & 2     & 2,579   & 37.3\%   & 0 \\
            CSS Conditional Rules Module, Level 3                         & CSS-CR & 1     & 449     & 36.5\%   & 0 \\
      CSS Font Loading Module, Level 3                              & CSS-FO & 12    & 2,560   & 33.5\%   & 0 \\
      CSS Object Model (CSSOM)                                      & CSS-OM & 15    & 8,193   & 12.6\%   & 0 \\
                  DOM, Level 1 - Specification                                  & DOM1   & 47    & 9,139   & 1.8\%    & 0 \\
      DOM, Level 2 - Core Specification                             & DOM2-C & 31    & 8,951   & 3.0\%    & 0 \\
      DOM, Level 2 - Events Specification                           & DOM2-E & 7     & 9,077   & 2.7\%    & 0 \\
      DOM, Level 2 - HTML Specification                             & DOM2-H & 11    & 9,003   & 4.5\%    & 0 \\
      DOM, Level 2 - Style Specification                            & DOM2-S & 19    & 8,835   & 4.3\%    & 0 \\
      DOM, Level 2 - Traversal and Range Specification              & DOM2-T & 36    & 4,590   & 33.4\%   & 0 \\
      DOM, Level 3 - Core Specification                             & DOM3-C & 10    & 8,495   & 3.9\%    & 0 \\
      DOM, Level 3 - XPath Specification                            & DOM3-X & 9     & 381     & 79.1\%   & 0 \\
      DOM Parsing and Serialization                                 & DOM-PS & 3     & 2,922   & 60.7\%   & 0 \\
                  execCommand                                                   & EC     & 12    & 2,730   & 24.0\%   & 0 \\
      File API                                                      & FA     & 9     & 1,991   & 58.0\%   & 0 \\
      Fullscreen API                                                & FULL   & 9     & 383     & 79.9\%   & 0 \\
      Geolocation API                                               & GEO    & 4     & 174     & 13.1\%   & 0 \\
                  HTML: Channel Messaging                                       & H-CM   & 4     & 5,018   & 77.4\%   & 0 \\
      HTML: Web Storage                                             & H-WS   & 8     & 7,875   & 29.2\%   & 0 \\
      HTML                                                          & HTML   & 195   & 8,980   & 4.3\%    & 0 \\
      HTML: History Interface                                       & H-HI   & 6     & 1,729   & 18.7\%   & 0 \\
                        Media Source Extensions                                       & MSE    & 8     & 1,616   & 37.5\%   & 0 \\
                        Performance Timeline                                          & PT     & 2     & 4,690   & 75.8\%   & 0 \\
      Performance Timeline, Level 2                                 & PT2    & 1     & 1,728   & 93.7\%   & 0 \\
                  Selection API                                                 & SEL    & 14    & 2,575   & 36.6\%   & 0 \\
      Selectors API, Level 1                                        & SLC    & 6     & 8,674   & 7.7\%    & 0 \\
                        Timing control for script-based animations                    & TC     & 1     & 3,568   & 76.9\%   & 0 \\
            UI Events Specification                                       & UIE    & 8     & 1,137   & 56.8\%   & 0 \\
            User Timing, Level 2                                          & UTL    & 4     & 3,325   & 33.7\%   & 0 \\
      DOM4                                                          & DOM4   & 3     & 5,747   & 37.6\%   & 0 \\
                  Non-Standard                                                  & NS     & 65    & 8,669   & 24.5\%   & 0 \\
    \bottomrule
  \end{tabular}
  \caption{Popularity and blockrate for the web standards that are used on at least 1\% of the Alexa 10k
or have at least one associated CVE advisory in the last three years.\\
\textbf{Columns one and two} list the name and abbreviation of the standard.\\
\textbf{Column three} gives the number of features (methods and properties) from that standard that we were able to instrument.\\
\textbf{Column four} includes the number of pages that used at least one feature from the standard, out of the entire Alexa 10k.\\
\textbf{Column five} shows the number of sites on which \textit{no features} in the standard executed in the presence of advertising and tracking blocking extensions (given that the website executed \textit{at least one feature} from the standard in the default case), divided by the number of pages where at least one feature from the standard was executed.  In other words, how often the blocking extensions prevented all features in a standard from executing, given at least one feature would have been used.\\
\textbf{Column six} shows the number of CVEs associated with this standard's implementation
in Firefox within the last three years.
}
  \label{fig:megatable}
\end{table*}
 
Just as all browser standards are not equally popular on the web, neither
are all standards equally associated with known vulnerabilities in \FF.  Some
standards have been associated with, or implicated in, a large number of
vulnerabilities, while others have not been associated with any publicly known
issues.  This subsection presents which browser standards have been connected to
known security vulnerabilities (in the form of filed CVEs), and the relative
popularity and block rates of those standards.

Column five of table~\ref{fig:megatable} shows the number of CVEs associated with this standard's implementation
in Firefox within the last three years.  As the table shows, some implementations of web standards have
been associated with a large number of security bugs even though those standards
are not popular on the web or are frequently blocked by advertising and
tracking blocking extensions.

For example, the \emph{Web Audio API}~\cite{webaudio2013standard} standard is an
example of the first case; a standard that is highly unpopular with website authors, but
which has exposed users to a substantial number of security vulnerabilities.
We observe the standard in use on fewer that 2\% of sites in our collection, but
its implementation in \FF is associated with at least 10 CVEs in the last 3 years.
\emph{WebRTC}~\cite{webrtcw3c} is used on less than 1\% of sites in the Alexa
10k, but is associated with 8 CVEs in the last 3 years.

The \emph{Scalable Vector Graphics}~\cite{svg2011standard} standard is
an example of the latter case.  The standard is very frequently blocked by
advertising and tracking blocking extensions; the standard is used on 1,554
sites in the Alexa 10k, but is prevented from executing in 87\% of cases.
At least 14 CVE's have been reported on \FF's implementation of the standard in
the last 3 years.

\subsection{Site Complexity}
\label{sec:results-site-complexity}

\begin{figure}
  \centering
  \includegraphics[width=\columnwidth]{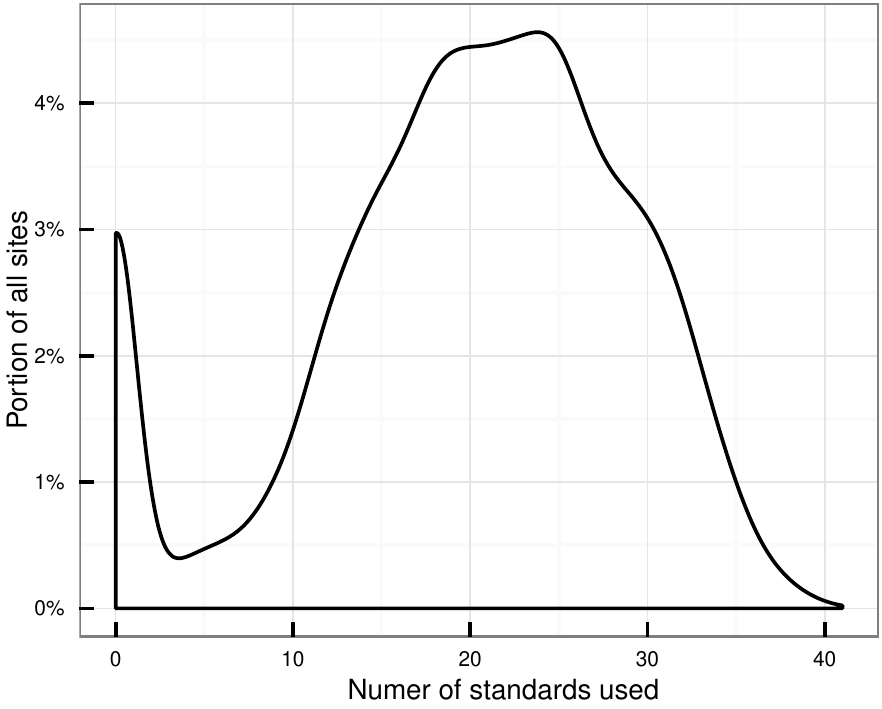}
  \caption{Probability density function of number of standards used by sites in the Alexa 10k.}
  \label{fig:results-site-complexity}
\end{figure}

Along with considering which standards are used by which sites, we can also
evaluate sites based on their complexity. We define complexity as the number of
standards used on a given website.  As
Figure~\ref{fig:results-site-complexity} shows, most sites use a reasonably
wide array of different standards: between 14 and 32 of the \numstandards
available in the browser. No site used more than 41 different standards, and a
second mode exists around the zero mark, showing that a small but measurable
subset of sites use little to no \JS at all.
 \section{Validation}
\label{sec:validation}
This study measures the features executed over repeated,
automated interactions with a website.  We then treat these
automated measurements as representative of the features that would be executed
when a human visited the website.

Our work relies on the assumption that our automated measurement technique
triggers (at least) all the browser functionality a human user's browser will
execute when interacting with the same website. This section explains how we
verified this assumption to be reasonable.

\subsection{Internal Validation}
\label{sec:internal-validation}

\begin{table}[H]
  \centering
  \begin{tabular}{ l r }
    \toprule
      Round \# &
      Avg. New Standards \\
    \midrule
      2 & 1.56 \\
      3 & 0.40 \\
      4 & 0.29 \\
      5 & 0.00 \\
    \bottomrule
  \end{tabular}
  \caption{Average number of new standards encountered on each subsequent automated crawl of a domain.}
  \label{fig:internal-validation-table}
\end{table}

As discussed in Section~\ref{sec:default-case-measurements}, we applied our automated
measurement technique to each site in the Alexa 10k five times.
We measured five times with the goal of capturing the full
set of functionality used on the site, since the measurement's
random walk technique means that each subsequent measurement may encounter
different parts of the site not reached previously.

A natural question then is whether five measurements are sufficient
to capture all potentially encountered features per site, or whether additional
measurements are necessary.  To ensure that five measurements where
sufficient, we examined how many new standards were encountered on
each round of measurement.  If new standards were still being encountered in the
final round of measurement, it would indicate we had not measured enough,
and that our data painted an incomplete picture of the types of features used
in each site.

Table \ref{fig:internal-validation-table} shows the results of this verification.
The first column lists each round of measurement, and the second
column lists the corresponding number of new standards encountered in the current
round that had not yet been observed in the previous rounds (averaged
across the entire Alexa 10k).  As the table shows, the average number of new
standards observed on each site decreased with each measurement,
until the 5th measurement of each site, at which point we did not observe any
new features being executed on any site.

From this we concluded that 5 rounds was a sufficient number of measurements for
each domain, and that further automated measurements of these sites were
unlikely to observe new feature usage.

\subsection{External Validation}
\label{sec:external-validation}

\begin{figure}[h]
  \centering
  \includegraphics[width=\columnwidth]{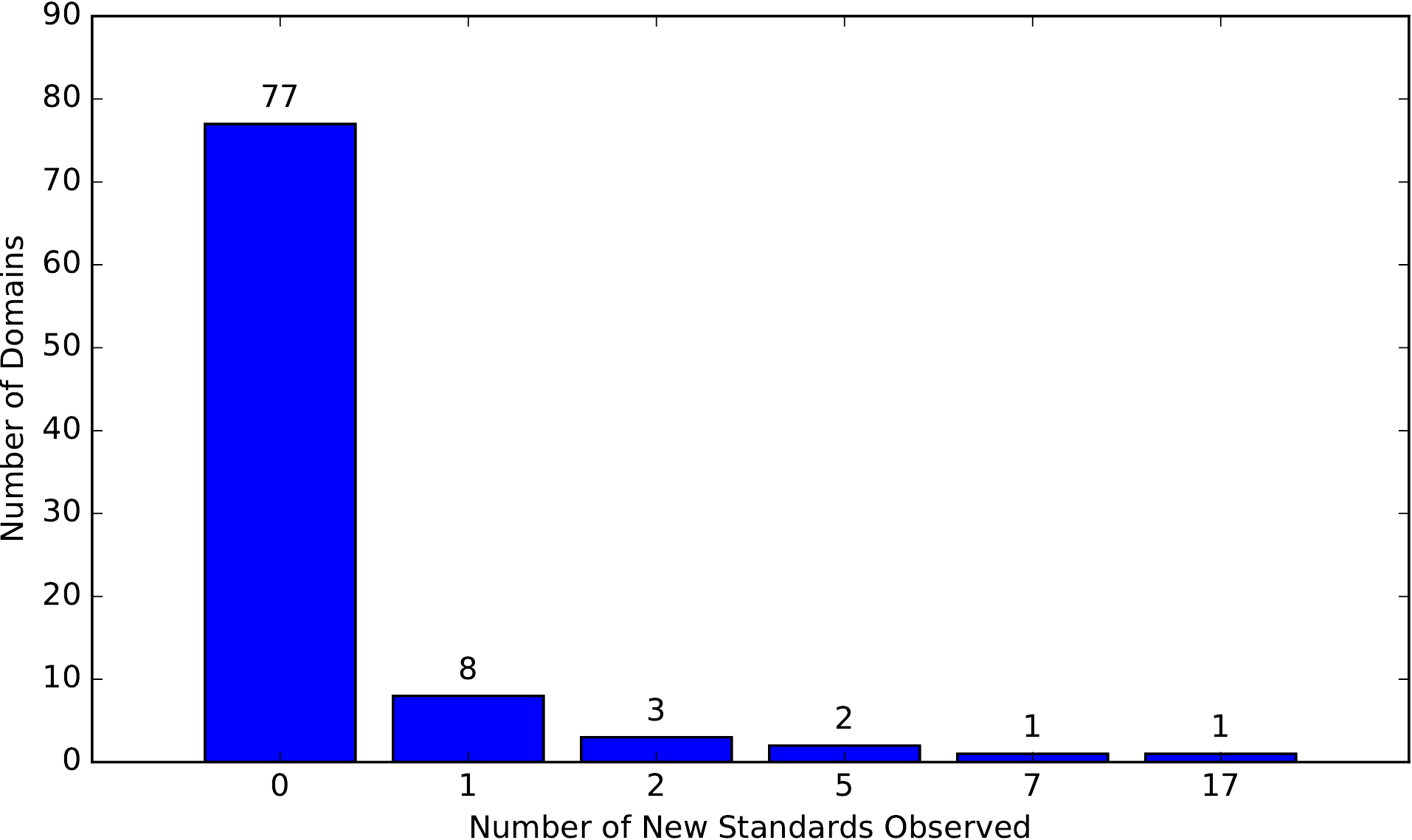}
  \caption{Average number of new standards encountered on each subsequent automated crawl of a domain.}
  \label{fig:external-validation-figure}
\end{figure}

We also tested whether our automated technique observes the same set of
functionalities as a human web user.   We chose 100 sites to visit randomly,
but weighted each choice according to the proportion of visits that site gets
according to Alexa. We continued to select sites until we had collected a set
of 100 different sites. We interacted with each site for 90 seconds in a casual
web browsing fashion.  This included reading articles and blog posts, scrolling
through websites, browsing site navigation listings, etc.  We interacted with
the home page of the site (the page directed to from the raw domain) for 30
seconds, then clicked on a prominent link we thought a typical human browser
would also choose (such as the headline of a featured article) and interacted
with this second page for 30 more seconds.  We then repeated the process a
third time, loading a third page that was interacted with for another 30
seconds.

After
omitting  pornographic and non-English sites,
We were able to complete this process for 92 different websites.
We then compared the features used during manual interaction with our automated
measurements of the same domains.
Figure \ref{fig:external-validation-figure} provides a histogram of this
comparison, with the x-axis showing the number of new standards
observed during manual interaction that \emph{were not} observed during
the automated interaction.  As the graph shows, in the majority of cases
(83.7\%), no new features were observed during manual interaction that the
automated measurements did not catch. The graph also shows a few outliers,
including one very significant outlier, where manual interaction triggered
standards that our automated technique did not.

From this we conclude that our automated measurement technique did a generally
accurate job of emulating the kinds of feature use a human user would
encounter on the web, even if the technique does not perfectly capture human
feature usage in all cases.

\vfill\eject
 \section{Discussion}
\label{sec:discussion}

In this section, we discuss the potential ramifications of these findings,
including what our results mean for the complexity of the browser.

\subsection{Popular and Unpopular Browser Features}

There are a small number of standards in the browser that are extremely
popular with website authors, providing features which can be thought of as
necessary for making modern web pages usable. These standards provide
functionality like querying the document for elements, inspecting and
validating forms, and making client-side page modifications.\footnote{All of
which are covered by the \emph{Document Object Model (DOM) Level 1
Specification} standard, dating back to 1998.}.

A much larger portion of the browser's functionality, however, is unused by most
site authors.  Eleven different \JS-exposed standards in \FF are completely
unused in the top ten thousand most popular websites, and 28 (nearly
37\% of standards available in the browser) are used by less than 1\% of sites.

While many unpopular features are relatively new to the browser, youth alone
does not seem to explain the extreme unpopularity of most features in the
browser on the open web. These lesser used features may be of interest only for those
creating applications which require login, or only small niches of developers
and site visitors.

\subsection{Blocked Browser Features}

When users employ common advertising and tracking blocking extensions, they
further reduce the frequency and number of standards that are executed in their
browser.  This suggests that some standards are primarily used to
support the advertising and tracking infrastructure built into the modern web.
When users browse with these common extensions installed, four additional standards
go unused on the web (a total of 15 standards, or 20\% of those available in the
browser).  An additional 20 standards become used on less than 1\% of websites
(for a total of 31 standards, or 41\% of standards in the browser).  16
standards are blocked over 75\% of the time by blocking extensions.

Furthermore,  while content blocker rules do not target \JS APIs
directly, that a standard like \emph{SVG}~\cite{svg2011standard},
used on 16\% of the Alexa 10k, would be prevented from running 87\% of
the time is circumstantial evidence that whatever website functionality this
enables is not necessary to the millions of people who use content blocking
extensions.  This phenomenon lends credence to what has been called ``the
Website Obesity Crisis'' - the conjecture that websites include far more
functionality than is actually necessary to serve the user's
purpose~\cite{ceglowski2015website}.

The presence of a large amount of unused functionality in the browser seems
to contradict the common security principal of least privilege, or of giving
applications only the capabilities they need to accomplish their intended
task, and no more.  This principal exists to limit attack surface and
limit the unforeseen security risks that can come from the unexpected,
and unintended, composition of features.  As the list of CVEs in
Figure~\ref{fig:megatable} shows, unpopular and heavily blocked features
have imposed substantial security costs to the browser.

Even though these features are frequently blocked, the sites
that they are blocked on are among the most popular websites in the
world. That these sites remain both functional and popular after having so much
of their functionality removed speaks to the robustness of the web programming
model, that these sites can still deliver the user's desired functionality even
after being heavily modified.

\subsection{Future Work}
This study develops and validates the use of monkey testing to elicit browser
feature usage on the open web. The closed web (i.e. web content and
functionality that are only available after logging in to a website) likely
uses a broader set of features. With the correct credentials, the monkey
testing approach could be used to evaluate those sites, although it would
likely need to be improved with an increased crawl depth or a rudimentary
understanding of site semantics.

Finally, a more complete treatment of the security implications of these broad
APIs would be valuable. In recent years, plugins like Java and
Flash have become less popular, and the native capabilities of browsers have
become more impressive.  The modern browser is a monolithic intermediary between
web applications and user hardware, like an operating system. For privacy
conscious users or those with special needs (like on public kiosks, or
electronic medical record readers), understanding the privacy and security
implications of this broad attack surface is very important.

 \section{Conclusion}

The Web API offers a standardized API for programming across operating systems
and web browsers.  This platform has been tremendously useful in the proliferation of the web as a
platform for both content dissemination and application distribution,
and has enabled the modern web, built on \JS and offering functionality like video, games, and
productivity applications.  Applications that were once only possible as
native apps or external plugins are now implemented in \JS in the browser.

With this move of the web from a content distribution system to an application platform,
more browser features have been added, not only support these applications,
but support them across the incredible range of devices we use to access the web.
Given this, it is not surprising that some features implemented within the browser are infrequently used.

Beyond this popularity divide, however, is the segment of features which
are blocked by content blockers in the vast majority of attempted uses.
Although consumers are not directly rejecting those features as complicit with ads or
tracking, the mere fact that these features are simultaneously popular with site authors
but overwhelmingly blocked by site users signals that these features may exist in the browser to
serve the needs of the site author rather than the site visitor.

That these features can even be blocked at all, however, speaks to the
robustness of the web's open standards and extensible user agents.  Preventing
such functionality in native applications is far less common and likely more
difficult. As the role of browser and the web continues to grow, the
ability of web users to customize their experience will likely remain an important
aspect of keeping the web user-centric, vibrant, and successful.
 
\bibliographystyle{acm}
{\small
\bibliography{references}
}

\end{document}